\documentclass[twocolumn]{aastex701}

\newcommand{\kms}{{km s$^{-1}$}}
\newcommand{\mas}{{mas yr$^{-1}$}}
\usepackage{hyperref}

\begin{document}

\title{Three Thousand Motion-Confirmed L and T Dwarf Candidates from the Backyard Worlds:~Planet 9 Citizen Science Project}

\author[0000-0002-6294-5937]{Adam C. Schneider}
\email{}
\affil{United States Naval Observatory, Flagstaff Station, 10391 West Naval Observatory Rd., Flagstaff, AZ 86005, USA}

\author[0000-0002-2387-5489]{Marc J. Kuchner}
\email{}
\affil{Exoplanets and Stellar Astrophysics Laboratory, NASA Goddard Space Flight Center, 8800 Greenbelt Road, Greenbelt, MD 20771, USA}

\author[0000-0001-6251-0573]{Jacqueline K. Faherty}
\email{}
\affil{Department of Astrophysics, American Museum of Natural History, Central Park West at 79th St., New York, NY 10024, USA}

\author[0000-0002-1125-7384]{Aaron M. Meisner}
\email{}
\affil{NSF National Optical-Infrared Astronomy Research Laboratory, 950 N. Cherry Ave., Tucson, AZ 85719, USA}
\affil{Center for Astrophysics $|$ Harvard \& Smithsonian, 60 Garden St., Cambridge, MA 02138, USA}
\affil{Radcliffe Institute for Advanced Study at Harvard University, 10 Garden Street, Cambridge, MA 02138, USA}

\author[0000-0003-4269-260X]{J. Davy Kirkpatrick}
\email{}
\affil{IPAC, Mail Code 100-22, Caltech, 1200 E. California Blvd., Pasadena, CA 91125, USA}

\author[0000-0002-6523-9536]{Adam J. Burgasser}
\email{}
\affil{Center for Astrophysics and Space Science, University of California San Diego, La Jolla, CA 92093, USA}

\author[0000-0001-8170-7072]{Daniella Bardalez Gagliuffi}
\email{}
\affil{Department of Physics \& Astronomy, Amherst College, 25 East Drive, Amherst, MA 01003, USA}

\author[0000-0003-2235-761X]{Thomas P. Bickle}
\email{}
\affil {School of Physical Sciences, The Open University, Milton Keynes, MK7 6AA, UK}

\author[0000-0001-7896-5791]{Dan Caselden}
\email{}
\affil{Department of Astrophysics, American Museum of Natural History, Central Park West at 79th St., New York, NY 10024, USA}

\author[0000-0003-2478-0120]{Sarah L. Casewell}
\email{}
\affil{School of Physics and Astronomy, University of Leicester, University Road, Leicester, LE1 7RH, UK}

\author[0000-0002-2592-9612]{Jonathan Gagn\'e}
\email{}
\affil{Plan\'etarium de Montr\'eal, Espace pour la Vie, 4801 av. Pierre-de Coubertin, Montr\'eal, Qu\'ebec, Canada}
\affil{Trottier Institute for Research on Exoplanets, Universit\'e de Montr\'eal, D\'epartement de Physique, C.P.~6128 Succ. Centre-ville, Montr\'eal, QC H3C~3J7, Canada}

\author[0000-0003-1202-3683]{Easton J. Honaker}
\email{}
\affil{Department of Physics and Astronomy, University of Delaware, Newark, DE 19716, USA}

\author[0000-0001-8662-1622]{Frank Kiwy}
\email{}
\affil{Backyard Worlds: Planet 9, USA}

\author[0000-0001-7519-1700]{Federico Marocco}
\email{}
\affil{IPAC, Mail Code 100-22, Caltech, 1200 E. California Blvd., Pasadena, CA 91125, USA}

\author[0000-0003-4083-9962]{Austin Rothermich}
\email{}
\affil{Department of Astrophysics, American Museum of Natural History, Central Park West at 79th St., New York, NY 10024, USA}

\author[0000-0003-4714-3829]{Nikolaj Stevnbak Andersen}
\email{}
\affil{Backyard Worlds: Planet 9, USA}

\author{Lizzeth Ruiz Arroyo}
\email{}
\affil{Backyard Worlds: Planet 9, USA}

\author[0000-0001-8731-9281]{Bruce Baller}
\email{}
\affil{Backyard Worlds: Planet 9, USA}

\author{Paul Beaulieu}
\email{}
\affil{Backyard Worlds: Planet 9, USA}

\author{John Bell}
\email{}
\affil{Backyard Worlds: Planet 9, USA}

\author[0009-0000-5790-7488]{Martin Bilsing}
\email{}
\affil{Backyard Worlds: Planet 9, USA}

\author{Troy K.~Bohling}
\email{}
\affil{Backyard Worlds: Planet 9, USA}

\author[0000-0002-7630-1243]{Guillaume Colin}
\email{}
\affil{Backyard Worlds: Planet 9, USA}

\author[0000-0002-8295-542X]{Giovanni Colombo}
\email{}
\affil{Backyard Worlds: Planet 9, USA}

\author[0009-0004-6814-5449]{Sam Deen}
\email{}
\affil{Backyard Worlds: Planet 9, USA}

\author{Alexandru Dereveanco}
\email{}
\affil{Backyard Worlds: Planet 9, USA}

\author{Kevin Dixon}
\email{}
\affil{Backyard Worlds: Planet 9, USA}

\author[0000-0002-4143-2550]{Hugo A. Durantini Luca}
\email{}
\affil{Backyard Worlds: Planet 9, USA}

\author[0009-0009-0264-1630]{Deiby Flores}
\email{}
\affil{Backyard Worlds: Planet 9, USA}

\author{Christoph Frank}
\email{}
\affil{Backyard Worlds: Planet 9, USA}

\author{Christopher Fulvi}
\email{}
\affil{Backyard Worlds: Planet 9, USA}

\author{Michael Gallmann}
\email{}
\affil{Backyard Worlds: Planet 9, USA}

\author[0000-0002-1044-1112]{Jean Marc Gantier}
\email{}
\affil{Backyard Worlds: Planet 9, USA}

\author{Konstantin Glebov}
\email{}
\affil{Backyard Worlds: Planet 9, USA}

\author[0000-0002-8960-4964]{L\'eopold Gramaize}
\email{}
\affil{Backyard Worlds: Planet 9, USA}

\author[0000-0002-7389-2092]{Leslie K. Hamlet}
\email{}
\affil{Backyard Worlds: Planet 9, USA}

\author[0000-0002-4733-4927]{Ken Hinckley}
\email{}
\affil{Backyard Worlds: Planet 9, USA}

\author{Kevin Jablonski}
\email{}
\affil{Backyard Worlds: Planet 9, USA}

\author[0000-0002-4175-295X]{Peter A. {Ja{\l}owiczor}}
\email{}
\affil{Backyard Worlds: Planet 9, USA}

\author[0000-0003-4905-1370]{Martin Kabatnik}
\email{}
\affil{Backyard Worlds: Planet 9, USA}

\author{Peter Kasprowitz}
\email{}
\affil{Backyard Worlds: Planet 9, USA}

\author[0009-0004-9268-9796]{K Ly}
\email{}
\affil{Backyard Worlds: Planet 9, USA}

\author{David W. Martin}
\email{}
\affil{Backyard Worlds: Planet 9, USA}

\author{Naoufel Marzak}
\email{}
\affil{Backyard Worlds: Planet 9, USA}

\author{Alexander McColgan}
\email{}
\affil{Backyard Worlds: Planet 9, USA}

\author{Neil J.~McEwan}
\email{}
\affil{Backyard Worlds: Planet 9, USA}

\author[0009-0000-8800-3174]{Marianne N. Michaels}
\email{}
\affil{Backyard Worlds: Planet 9, USA}

\author{William Pendrill}
\email{}
\affil{Backyard Worlds: Planet 9, USA}

\author{St{\'e}phane Perlin}
\email{}
\affil{Backyard Worlds: Planet 9, USA}

\author[0000-0001-9692-7908]{Ben Pumphrey}
\email{}
\affil{Backyard Worlds: Planet 9, USA}

\author{James Rabe}
\email{}
\affil{Backyard Worlds: Planet 9, USA}

\author{Henry Raway}
\email{}
\affil{Backyard Worlds: Planet 9, USA}

\author{Walter Ruben Robledo}
\email{}
\affil{Backyard Worlds: Planet 9, USA}

\author{David Roser}
\email{}
\affil{Backyard Worlds: Planet 9, USA}

\author[0009-0005-4611-4008]{Animesh Roy}
\email{}
\affil{Backyard Worlds: Planet 9, USA}
\affil{Rajshahi University of Engineering \& Technology, Kazla, Rajshahi-6204, Bangladesh}

\author[0000-0003-4864-5484]{Arttu Sainio}
\email{}
\affil{Backyard Worlds: Planet 9, USA}

\author[0009-0000-6624-8031]{Vincent Schindler}
\email{}
\affil{Backyard Worlds: Planet 9, USA}

\author{Manfred Schonau}
\email{}
\affil{Backyard Worlds: Planet 9, USA}

\author[0000-0002-7587-7195]{J{\"o}rg Sch{\"u}mann}
\email{}
\affil{Backyard Worlds: Planet 9, USA}

\author{Karl Selg-Mann}
\email{}
\affil{Backyard Worlds: Planet 9, USA}

\author{Andrea Serio}
\email{}
\affil{Backyard Worlds: Planet 9, USA}

\author[0009-0005-3530-9571]{David Sirbescu-Stanley}
\email{}
\affil{Backyard Worlds: Planet 9, USA}
\affil{Western Michigan University, 1903 W. Michigan Ave, Kalamazoo, MI 49008, USA}

\author{Patrick Smith}
\email{}
\affil{Backyard Worlds: Planet 9, USA}

\author{Andres Stenner}
\email{}
\affil{Backyard Worlds: Planet 9, USA}

\author{Christine Sunjoto}
\email{}
\affil{Backyard Worlds: Planet 9, USA}

\author{Christopher Tanner}
\email{}
\affil{Backyard Worlds: Planet 9, USA}

\author[0000-0001-5284-9231]{Melina Th{\'e}venot}
\email{}
\affil{Backyard Worlds: Planet 9, USA}

\author{Vinod Thakur}
\email{}
\affil{Backyard Worlds: Planet 9, USA}

\author[0000-0002-3878-7166]{Mayahuel Torres Guerrero}
\email{}
\affil{Backyard Worlds: Planet 9, USA}

\author{Maurizio Ventura}
\email{}
\affil{Backyard Worlds: Planet 9, USA}

\author{Nikita V. Voloshin}
\email{}
\affil{Backyard Worlds: Planet 9, USA}

\author{Jim Walla}
\email{}
\affil{Backyard Worlds: Planet 9, USA}

\author{Zbigniew W\c edracki}
\email{}
\affil{Backyard Worlds: Planet 9, USA}

\author{Bailey Weyandt}
\email{}
\affil{Backyard Worlds: Planet 9, USA}

\author{Breck Wilhite}
\email{}
\affil{Backyard Worlds: Planet 9, USA}

\author{Spartacus Zitouni}
\email{}
\affil{Backyard Worlds: Planet 9, USA}

\begin{abstract}
The Backyard Worlds: Planet 9 citizen science project uses data from the Wide-field Infrared Survey Explorer to detect infrared objects with significant motion.  In this work, we present the majority of the L and T dwarf candidates discovered through this effort.  For each candidate, we provide proper motion measurements as well as optical, near-infrared, and mid-infrared photometry (when available), photometric spectral types and distance estimates.  Three thousand and six new motion-confirmed discoveries are presented in this work, 2,357 with L-type photometric spectral types and 649 with T-type photometric spectral types.  We also present an additional 80 objects as likely L or T dwarfs based on available photometry, but for which a significant motion measurement could not be obtained. We identify 28 objects in this sample as new comoving companions to higher-mass stars, and an additional 9 sources that are candidate binary systems made up of two ultracool dwarfs of L-type or later.  Follow-up spectroscopic observations will be necessary to confirm spectral types and further characterize the sources discovered through this project.  This work presents the largest single sample of motion-confirmed L and T dwarf discoveries to date, which would more than double the number of known L and T dwarfs, if confirmed.  We wish to sincerely thank our citizen scientist collaborators for their monumental efforts that have directly impacted this project's success.
\end{abstract}

\section{Introduction} 
\label{sec:intro}

The collective population of brown dwarfs identified since their discovery \citep{nakajima1995, rebolo1995} has allowed for significant advances in numerous fields of astrophysics, from a more detailed understanding of how molecular clouds collapse into stars and brown dwarfs (e.g., \citealt{kirkpatrick2012, kirkpatrick2019, kirkpatrick2021, kirkpatrick2024, gagne2017, lodieu2019, best2024}), to the low-mass star formation history of the Milky Way through the analysis of the oldest, most metal-poor brown dwarfs (e.g., \citealt{burgasser2003b, burgasser2025, burgasser2025b, lepine2003, burgasser2006b, cushing2009, mace2013b, burningham2014, kirkpatrick2014, kirkpatrick2021b, zhang2017, zhang2017b, zhang2018a, zhang2018b, zhang2019a, zhang2019b, gonzales2018, kellogg2018, greco2019, zhang2019c, zhang2024b, gonzales2020, schneider2020, meisner2021, meisner2023b, lodieu2022, zhang2023b, faherty2025}), to helping provide a more comprehensive view of giant exoplanet atmospheres (e.g., \citealt{faherty2013, faherty2016, liu2016, gagne2017b, miles2018, miles2023, vos2022, gonzalez2024, kiman2026}).  L and T dwarfs are typically found using their unique optical and infrared colors (e.g., \citealt{burgasser1999, burgasser2002, burgasser2004, kirkpatrick1999, kirkpatrick2000, kirkpatrick2008, kirkpatrick2011, gizis2000, hawley2002, cruz2003, cruz2007, knapp2004, chiu2006, reid2008a, west2008, zhang2009, burningham2010, burningham2013, schmidt2010, albert2011, cushing2011, lodieu2012, best2013, best2015, dayjones2013, mace2013, thompson2013, cardoso2015, gagne2015b, marocco2015, skrzypek2016, kellogg2017, schneider2017, reyle2018, tinney2018, carnero2019, kiman2019, sorahana2019, zhang2024, dominguez2025, zerjal2025, kiwy2026}) or significant proper motions (e.g., \citealt{kirkpatrick2010, kirkpatrick2014, kirkpatrick2016, deacon2014, luhman2014, smith2014, aller2016, robert2016, schneider2016, theissen2017, smith2018, greco2019, meisner2020a, karpov2025}).  Despite numerous previous surveys for such objects, a complete accounting of them in existing wide-area survey data has been difficult to achieve. 

The Backyard Worlds: Planet 9 (BYW) citizen science project (\href{http://www.backyardworlds.org/}{http://www.backyardworlds.org/}) was designed to use data from the Wide-field Infrared Survey Explorer (WISE; \citealt{wright2010}) to identify objects with significant motion across the sky, with a focus on substellar objects and distant planets in our solar system \citep{kuchner2017}.  The efforts of the thousands of volunteers who have engaged in the project have led to numerous discoveries, including very low-mass binaries \citep{faherty2020, softich2022, bravo2023, humphreys2023}, new members of the Y spectral class \citep{bardalez2020, meisner2020, robbins2023}, low-metallicity subdwarfs \citep{schneider2020, meisner2021, brooks2022, burgasser2024, burgasser2025}, co-moving companions (\citealt{kiwy2021, rothermich2021, rothermich2024, schneider2021, faherty2021, gramaize2022, gramaize2024, kiwy2022, marocco2024, jalowiczor2026}), young objects (\citealt{robbins2023a, schneider2023, bickle2024, hyogo2025}; Bickle et al.~in prep.), unique white dwarf systems (\citealt{debes2019, jalowiczor2021b, bickle2022, bravo2025}; Casewell et al.~in prep.), as well as numerous other high proper motion, nearby, and/or very cold sources (\citealt{kirkpatrick2021, kirkpatrick2024, kota2022, schapera2022, brooks2024, wood2025, zhang2025, kiwy2026}; Raghu et al.~in prep.; Honaker et al.~in prep.). 

In this work, we present the vast majority of L and T dwarf candidates found through the BYW project. These objects are prime targets for additional characterization from pointed follow-up observations as well as ongoing and upcoming surveys such as the Rubin Observatory Legacy Survey of Space and Time (LSST; \citealt{ivezic2019}), the Nancy Grace Roman Space Telescope (Roman; \citealt{spergel2015}), the Euclid Space Telescope (Euclid; \citealt{laureijs2011, scaramella2022}), and the Spectro-Photometer for the History of the Universe, Epoch of Reionization, and Ices Explorer (SPHEREx; \citealt{dore2014}).

In Section \ref{sec:candidates} we describe the candidate selection process for objects to be included in this compilation.  In Section \ref{sec:properties} we describe the general properties of the sample, and in Section \ref{sec:objects} we highlight some of the more unique discoveries, with a focus on systems with multiple components. In Section \ref{sec:conclusion} we conclude with a summary of this work and recommendations for future efforts to fully characterize this new compilation of objects.

\section{Candidate Selection} 
\label{sec:candidates}

The BYW project asks volunteers to identify objects with significant motions in WISE data.  Volunteers are asked to inspect ``flipbooks'' of time-resolved WISE images \citep{Meisner_2018}, processed to highlight moving sources.  However, several enterprising volunteers have developed their own methods of identifying ultracool dwarf candidates through various catalog queries and image analyses.  While the use of multiple methods leads to a somewhat heterogeneous sample, making characteristics such as survey completeness difficult to establish, attacking the problem of ultracool dwarf discovery from multiple angles has unveiled a diverse array of discoveries. Regardless of the method of discovery, the project provides an online submission form called the ``Think-You've-Got-One'' (or TYGO) form for submitting individual objects for review by the science team.  As of November 2025, the TYGO form has over 102,000 individual submissions. 

Each of these submissions is individually inspected by members of the science team to determine whether or not the particular submitted object warrants inclusion in our list of candidates that are suitable for future follow-up observations.  The primary requirements are 1) discernible motion and 2) a photometric spectral type estimate of L0 or later.  For many objects, motion can be seen by-eye using multi-epoch WISE images via the WiseView tool \citep{caselden2018}.  For some sources, additional steps are taken to confirm significant motion through existing proper motion catalogs or measuring motion via individual detections, as described in Section \ref{sec:pms}.

Spectral type estimates are determined by first gathering available archival optical, near-infrared, and/or mid-infrared photometry for each moving source.  This includes optical photometry from Gaia (Gaia DR3; \citealt{gaia2023}), the Dark Energy Survey (DES; \citealt{abbott2018, abbott2021}), the Panoramic Survey Telescope and Rapid Response System (Pan-STARRS; \citealt{chambers2016, flewelling2020}), and/or the NOIRLab Source Catalog (NSC; \citealt{nidever2021}). Note that for optical photometry, we do not include limits or measurements with uncertainties $\geq$ 1 mag.  Near-infrared photometry is collected from the 2-Micron All Sky-Survey (2MASS; \citealt{skrutskie2006}), UKIRT Infrared Deep Sky Survey (UKIDSS; \citealt{lawrence2007, lucas2008}), the UKIRT Hemisphere Survey (UHS; \citealt{dye2018, schneider2025}), the VISTA Hemisphere Survey (VHS; \citealt{mcmahon2013}), VISTA Variables in the V{\'i}a L{\'a}ctea (VVV; \citealt{minniti2010, saito2012}), and the VISTA Kilo-degree Infrared Galaxy Survey (VIKING; \citealt{edge2013}) surveys.  Finally, CatWISE2020 \citep{marocco2021} photometry is gathered for every object.

We then estimate spectral types for each source using the procedure described in \cite{schneider2016}.  This method uses available color combinations compared to a training set of the colors of known objects using a k-Nearest Neighbors machine learning algorithm. For this work, we do not include any objects with ``M'' spectral type estimates.  Those found to be possible L, T, or Y dwarfs are retained for further analysis.  

Over 3,000 as yet unpublished candidate L and T dwarfs were found through this effort, as described in Section \ref{sec:properties}.  

\startlongtable
\begin{deluxetable*}{llccccccc}
\label{tab:props}
\tablecaption{BYW Motion Discovery Properties}
\tablehead{
\colhead{Column Label} & \colhead{Description} & \colhead{Example} & Units }
\startdata
Name-CWISE & CatWISE2020 designation & J000002.09+443725.2 & \dots \\
pmRA & Proper motion in RA & -30.5 & mas yr$^{-1}$ \\
epmRA & Uncertainty of proper motion in RA & 2.1 & mas yr$^{-1}$ \\
pmDE & Proper motion in Dec. & 52.6 & mas yr$^{-1}$ \\ 
epmDE & Uncertainty of proper motion in Dec. & 2.1 & mas yr$^{-1}$ \\
pm-source & source of the measured proper motion & 1 & \dots \\
Gmag & Gaia G magnitude & \dots & mag \\
eGmag & Uncertainty of Gaia G magnitude & \dots & mag \\
BPmag & Gaia G$_{\rm BP}$ magnitude & \dots & mag \\
eBPmag & Uncertainty of Gaia G$_{\rm BP}$ magnitude & \dots & mag \\
RPmag & Gaia G$_{\rm RP}$ magnitude & \dots & mag \\
eRPmag & Uncertainty of Gaia G$_{\rm RP}$ magnitude & \dots & mag \\
gmag & $g$ magnitude & \dots & mag \\
egmag & Uncertainty of $g$ magnitude & \dots & mag \\
rmag & $r$ magnitude & \dots & mag \\
ermag & Uncertainty of $r$ magnitude & \dots & mag \\
imag & $i$ magnitude & \dots & mag \\
eimag & Uncertainty of $i$ magnitude & \dots & mag \\
zmag & $z$ magnitude & 20.389 & mag \\
ezmag & Uncertainty of $z$ magnitude & 0.039 & mag \\
ymag & $y$ magnitude & 19.357 & mag \\
eymag & Uncertainty of $y$ magnitude & 0.035 & mag \\
opt-source & Source of optical photometry & 21 & \dots \\
Ymag & $Y$ magnitude & \dots & mag \\
eYmag & Uncertainty of $Y$ magnitude & \dots & mag \\
Jmag & $J$ magnitude & 17.138 & mag \\
eJmag & Uncertainty of $J$ magnitude & 0.024 & mag \\
Hmag & $H$ magnitude & 16.309 & mag \\
eHmag & Uncertainty of $H$ magnitude & 0.020 & mag \\
Kmag & $K$ magnitude & 15.551 & mag \\
eKmag & Uncertainty of $K$ magnitude & 0.023 & mag \\
ir-source & Source of near-IR photometry & 1 & \dots \\
W1mag & CatWISE2020 W1 magnitude & 14.988 & mag \\
eW1mag & Uncertainty of CatWISE2020 W1 magnitude & 0.017 & mag \\
W2mag & CatWISE2020 W2 magnitude & 14.811 & mag \\
eW2mag & Uncertainty of CatWISE2020 W2 magnitude & 0.025 & mag \\
BYW-discoverer-code\tablenotemark{a} & BYW discoverer code & A & \dots \\
phot-type & Photometric spectral type & L4 & \dots \\
J-dist & $J$-band distance estimate & 76 & pc \\
W2-dist & W2 distance  estimate & 69 & pc \\
Note & Note\tablenotemark{b} & \dots & \dots \\
\enddata
\tablerefs{(1) UHS DR3 \citep{schneider2025}; (2) unTimely detections \citep{meisner2023} (this work); (3) VHS detections \citep{mcmahon2013} (this work); (4) Pan-STARRS DR2 detections \citep{chambers2016, flewelling2020} (this work); (5) UHS detections \citep{dye2018} (this work); (6) UKIDSS LAS \citep{lawrence2007}; (7) NSC DR2 \citep{nidever2021}; (8) Gaia DR3 \citep{gaia2023}; (9) NSC DR2 detections \citep{nidever2021} (this work); (10) VIKING detections \citep{edge2013} (this work); (11) CatWISE2020 \citep{marocco2021}; (12) VMC detections \citep{cioni2011} (this work); (13) UKIDSS LAS detections \citep{lawrence2007} (this work); (14) UKIDSS GCS \citep{lawrence2007}; (15) UKIDSS GCS detections \citep{lawrence2007} (this work); (16) UKIDSS GPS detections \citep{lucas2008} (this work); (17) UKIDSS GPS \citep{lucas2008}; (18) UKIDSS DXS \citep{lawrence2007}; (19) VIRAC2 \citep{smith2025}; (20) VVV detections \citep{minniti2010} (this work); (21) Pan-STARRS DR2 \citep{chambers2016, flewelling2020}; (22) DES DR2 \citep{abbott2021}; (23) VHS \citep{mcmahon2013}; (24) 2MASS \citep{skrutskie2006}; (25) 2MASS reject catalog \citep{skrutskie2006}; (26) VIKING \citep{edge2013}; (27) VMC \citep{cioni2011}; (28) VVV \citep{minniti2010}  }
\tablecomments{This table is available in its entirety in a machine-readable form in the online journal.}
\tablenotetext{a}{BYW Discoverer Codes:
A = William Pendrill,
B = Sam Deen,
C = Christopher Tanner,
D = Arttu Sainio,
E = Dan Caselden,
F = Martin Kabatnik, 
G = Peter Kasprowitz,
H = Bruce Baller, 
I = Tom Bickle,
J = Paul Beaulieu, 
K = Sam Goodman, 
L = Patrick Smith,
M = Animesh Roy,
N = Frank Kiwy, 
O = Naoufel Marzak, 
P = Mayahuel Torres Guerrero,
Q = Jean Marc Gantier, 
R = Karl Selg-Mann, 
S = Nikolaj Stevnbak Andersen, 
T = David W.~Martin, 
U = Nikita Valerievich Voloshin, 
V = L{\'e}opold Gramaize, 
W = Les Hamlet, 
X = Austin Rothermich, 
Y = Alexandru Dereveanco, 
Z = Zbigniew W{\c{e}}dracki,
AA = Melina Th{\'e}venot, 
BB = Guillaume Colin, 
CC = rhocagil,
DD = Andres Guillermo Stenner, 
EE = Troy K. Bohling,
FF = J{\"o}rg Sch{\"u}mann, 
GG = Maurizio Ventura, 
HH = James Rabe, 
II = Andrea Serio,
JJ = Marianne N.~Michaels, 
KK = David Roser, 
LL = Walter Ruben Robledo, 
MM = Vinod Thakur, 
NN = Deiby Flores, 
OO = Michael Gallmann, 
PP = Lizzeth Ruiz Arroyo,
QQ = Ken Hinckley, 
RR = Giovanni Colombo, 
SS = K Ly,
TT = Kevin Dixon, 
UU = Jim Walla,
VV = Neil J.~McEwan,
WW = Christine Sunjoto,
XX = Ben Pumphrey,
YY = Martin Bilsing,
ZZ = Hugo A.~Durantini Luca,
AAA = Spartacus Zitouni,
BBB = Breck Wilhite,
CCC = Bailey Weyandt,
DDD = Henry Raway,
EEE = John Bell,
FFF = Manfred Schonau,
GGG = Alexander McColgan,
HHH = Vincent Schindler, 
III = Christopher Fulvi,
JJJ = Kevin Jablonski,
LLL = David Sirbescu-Stanley,
MMM = Konstantin Glebov,
NNN = Peter Ja{\l}owiczor}
\tablenotetext{b}{Notes: 1 = WISE data from CatWISE2020 Reject Catalog; 2 = this object has proper motion significance less than 3$\sigma$ }
\end{deluxetable*}

\section{Sample Properties} 
\label{sec:properties}

The entire sample of newly discovered L and T dwarf candidates can be found in Table \ref{tab:props}, which also includes the names of the BYW volunteers who helped in the discovery of each object.

\subsection{Proper Motions}
\label{sec:pms}

Ultimately, for this work, we keep as candidates all objects for which we were able to measure a total proper motion significance of $>$3$\sigma$.  Proper motions come from several sources, most notably CatWISE2020, Gaia DR3, UKIDSS, UHS, and NSC DR2.  For some sources, we were able to calculate more precise proper motions than those in existing databases by using the availability of individual detections from specific catalogs, including unTimely \citep{meisner2023}, UKIDSS, VHS, VVV, and VIKING, DES, Pan-STARRS, and NSC DR2.  We did not mix detections from different catalogs to determine proper motions, with the exception of UKIDSS and/or VISTA survey detections as these are processed using the same astrometric pipeline. Proper motions calculated in this work using individual detections were determined by using a least-squares fit weighted by the quoted uncertainty of each astrometric measurement. If more than one proper motion exists for a source, we use the measurement with the highest precision (e.g., smallest uncertainties).  Proper motions for each candidate are given in Table \ref{tab:props}. A future effort to re-register all previously mentioned catalogs to a common reference frame (e.g., \citealt{gaia2016, gaia2023}) would certainly increase the precision of many of the proper motions included in Table \ref{tab:props}, but is beyond the scope of this work.  

A typical proper motion for our new discoveries is $\sim$150 \mas, with the median and average of the total proper motion distribution being 141 \mas\ and 161 \mas, respectively.  Ninety percent of the proper motions found for the objects in this work are between 60 \mas and 325 \mas.  Five objects were found with total proper motion values greater than 1000 \mas\ (CWISE J043407.35$-$233132.9, CWISE J084405.60$-$105250.4, CWISE J123908.72+073226.1, CWISE J133928.30$-$513546.8, CWISE J205435.40+582143.6), with the largest motion belonging to CWISE J133928.30$-$513546.8 at 1882$\pm$174 \mas.

An additional 80 objects are included that have colors and spectral type estimates indicative of L or T dwarfs, but for which we were not able to determine a significant proper motion.  The majority of these sources are faint candidates with large positional uncertainties, resulting in large proper motion uncertainties.  These candidates are flagged as having low proper motion significance in Table \ref{tab:props}.

\subsection{Spectral Type Estimates}
\label{sec:spts}
\begin{figure}
\plotone{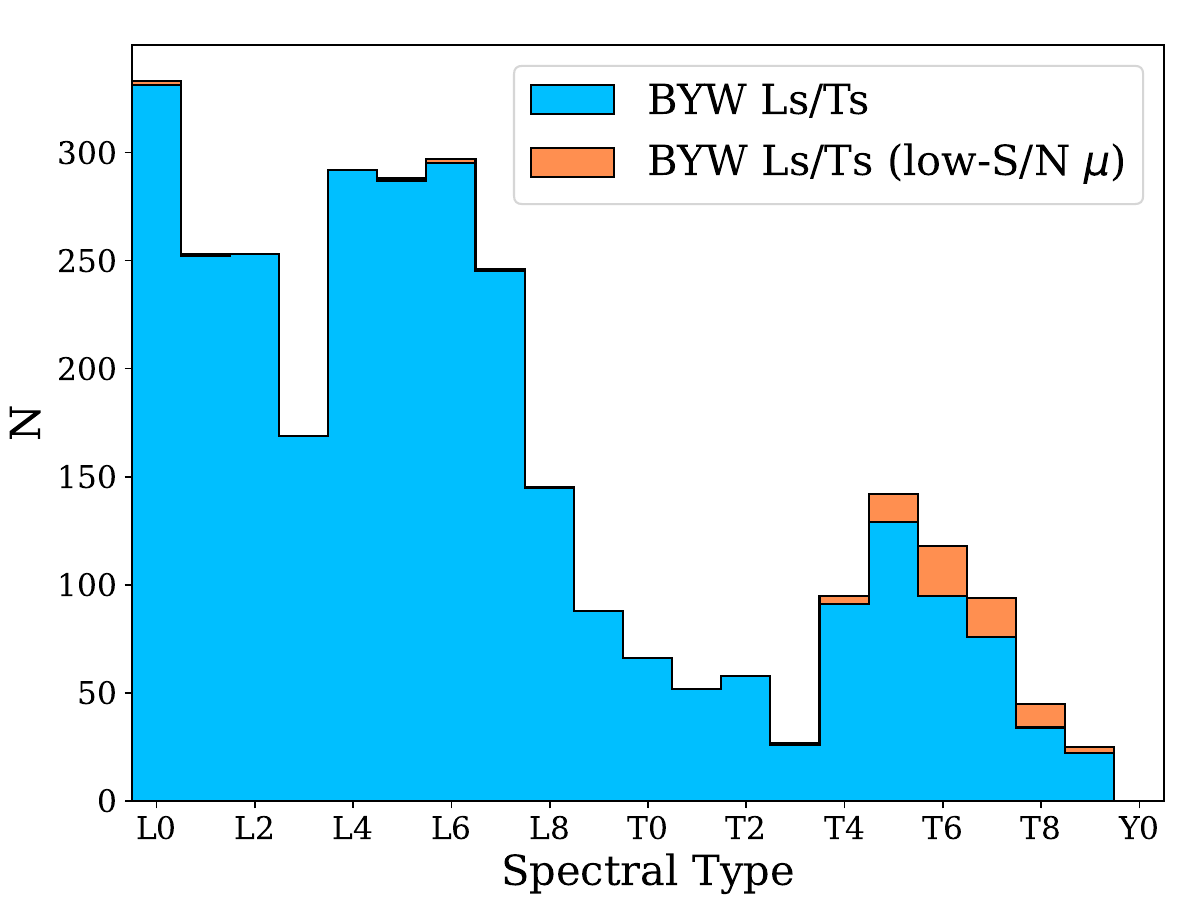}
\caption{Histogram of spectral type estimates for the BYW sample.  The orange histogram shows the additional objects that are likely new L or T dwarf discoveries but for which significant motion could not be confirmed. }  
\label{fig:spthist}
\end{figure}

As discussed in Section \ref{sec:candidates}, spectral type estimates were determined using the method in the appendix of \cite{schneider2016}, which has generally been found to be accurate to within $\pm$1 subtypes for field L and T dwarfs.  Of the 3,006 new discoveries presented in this work, 2,357 have L-type photometric spectral type estimates and 649 have T-type photometric spectral type estimates.  The majority of the Y dwarf candidates found through the BYW project have been published previously \citep{bardalez2020, meisner2020, kirkpatrick2021, robbins2023, kirkpatrick2024}, so few strong Y dwarf candidates are presented in this work.  There are 28 WISE W2-only objects for which color limits only allow us to place limits on spectral type estimates.  Some of these objects could be Y dwarfs, but more information is needed in order to confidently assign spectral types. Table \ref{tab:w2s} lists these 28 W2-only sources and their W1-W2 color limits.  A histogram showing the distribution of spectral type estimates for the entire sample is shown in Figure \ref{fig:spthist}.  

\begin{deluxetable}{lccccccc}
\tabletypesize{\scriptsize}
\label{tab:w2s}
\tablecaption{W2-only Sources}
\tablehead{
\colhead{CatWISE} & \colhead{W1} & \colhead{W2} & \colhead{W1-W2} \\
\colhead{Name} & \colhead{(mag)} & \colhead{(mag)} & \colhead{(mag)}}
\startdata
J002032.37+872303.7	&	$>$20.028	&	16.723	$\pm$	0.082	&	$>$3.305	\\
J002622.96$-$163129.7	&	$>$19.609	&	16.349	$\pm$	0.092	&	$>$3.260	\\
J010951.45+061301.3\tablenotemark{a}	&	$>$19.600	&	16.070	$\pm$	0.092	&	$>$3.530	\\
J011840.00-385342.5\tablenotemark{a}	&	$>$19.874	&	16.312	$\pm$	0.073	&	$>$3.562	\\
J014622.57+040639.2\tablenotemark{a}	&	$>$19.088	&	16.279	$\pm$	0.087	&	$>$2.809	\\
J024615.03$-$462648.5	&	$>$19.933	&	16.590	$\pm$	0.087	&	$>$3.343	\\
J031917.56+522109.2	&	$>$19.356	&	16.333	$\pm$	0.105	&	$>$3.023	\\
J031957.40+121222.7	&	$>$19.508	&	16.173	$\pm$	0.087	&	$>$3.335	\\
J040434.09+252747.0\tablenotemark{a}	&	$>$19.201	&	16.126	$\pm$	0.076	&	$>$3.075	\\
J045533.14+024658.8	&	$>$19.314	&	16.606	$\pm$	0.113	&	$>$2.708	\\
J065229.89$-$595729.9\tablenotemark{a}	&	$>$19.549	&	16.758	$\pm$	0.070	&	$>$2.791	\\
J095307.80+171807.1	&	$>$19.580	&	16.301	$\pm$	0.086	&	$>$3.279	\\
J102420.32$-$045647.7	&	$>$19.593	&	16.321	$\pm$	0.100	&	$>$3.272	\\
J112404.78+264109.3	&	$>$19.650	&	16.400	$\pm$	0.087	&	$>$3.250	\\
J130529.87+754856.7	&	$>$20.067	&	16.758	$\pm$	0.087	&	$>$3.309	\\
J133303.32+195354.2	&	$>$19.246	&	16.407	$\pm$	0.088	&	$>$2.839	\\
J135606.62+262609.1\tablenotemark{a}	&	$>$19.179	&	16.705	$\pm$	0.110	&	$>$2.474	\\
J141727.53$-$021328.8	&	$>$19.367	&	16.140	$\pm$	0.079	&	$>$3.227	\\
J162509.38$-$102829.6	&	$>$18.438	&	15.638	$\pm$	0.069	&	$>$2.800	\\
J165935.77+753821.4\tablenotemark{a}	&	$>$19.917	&	16.680	$\pm$	0.078	&	$>$3.237	\\
J180008.29$-$472225.2	&	$>$18.815	&	15.888	$\pm$	0.058	&	$>$2.927	\\
J181011.19+654319.0	&	$>$20.543	&	16.924	$\pm$	0.037	&	$>$3.619	\\
J194046.45+804155.2	&	$>$19.942	&	16.311	$\pm$	0.055	&	$>$3.631	\\
J194325.28$-$212243.1	&	$>$19.344	&	16.230	$\pm$	0.097	&	$>$3.114	\\
J195303.00+770413.4	&	$>$20.021	&	16.649	$\pm$	0.072	&	$>$3.372	\\
J205435.40+582143.6	&	$>$19.933	&	16.837	$\pm$	0.101	&	$>$3.096	\\
J232053.14$-$820932.7	&	$>$19.343	&	16.663	$\pm$	0.082	&	$>$2.680	\\
J232218.67$-$790754.9	&	$>$19.963	&	16.744	$\pm$	0.087	&	$>$3.219	
\enddata
\tablenotetext{a}{Low proper motion significance.}
\end{deluxetable}

Because our search for new L and T dwarf candidates is largely WISE magnitude limited, it is not unexpected that many of the discoveries are of early-L type, as these are much brighter than later-type brown dwarfs at a similar distance.  There are three notable dips in the spectral type distribution of our new discoveries; around spectral type L3, between $\sim$T0 and T3, and for types later than T8.  The underlying cause of the dip around L3 is unclear, but could be due to the limiting reach of Gaia for identifying early L dwarfs (e.g., \citealt{reyle2018, smart2021, sarro2023}) intersecting with the limit of our project's survey to detect such objects at WISE limiting magnitudes.  The dearth of objects at early T types likely coincides with the rapid evolution of objects just beyond the L/T transition, as noted in \cite{best2021}.  The lack of objects with spectral types beyond T8 can likely be attributed to two factors, the first being that most of the latest spectral type candidates found through the BYW project have been published previously (most notably in the 20 pc sample compilations of \citealt{kirkpatrick2021} and \citealt{kirkpatrick2024}), and the second is that cold T9 and later type objects are inherently very faint, so our survey has searched a much smaller volume of space for these objects compared to earlier types.

\begin{figure*}
\plotone{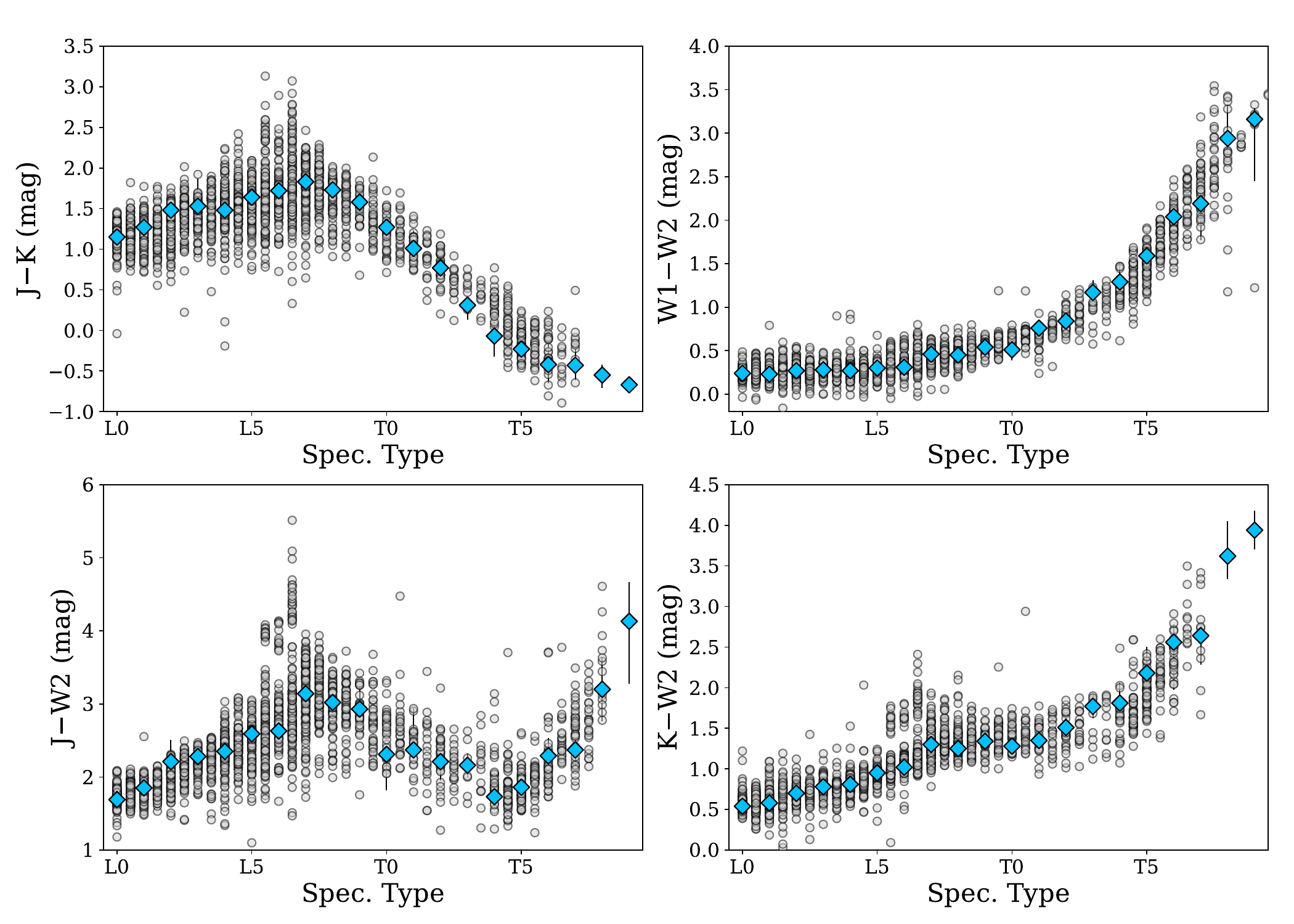}
\caption{Colors as a function of estimated spectral type for this sample (gray symbols) compared to median colors from the sample of \cite{schneider2023} (blue diamonds).  Newly discovered objects generally follow the trends of known objects, with outliers typically due to unusual colors or blending in WISE images. }  
\label{fig:colors}
\end{figure*}

Figure \ref{fig:colors} shows infrared colors ($J-K$, $J-$W2, $K-$W2, and W1$-$W2) as a function of spectral type for the entire sample.  Also shown are median values for known L and T dwarfs from the sample of \cite{schneider2023} where outliers (e.g, young objects, subdwarfs, binaries) are excluded.  The figure shows that the majority of objects have colors that follow the expected trends with spectral type.  Some outliers may be objects with unusual colors for their predicted spectral type, however many of these outliers are due to blended or uncertain photometry.

\subsection{Distances}
\label{sec:dist}

\begin{figure}
\plotone{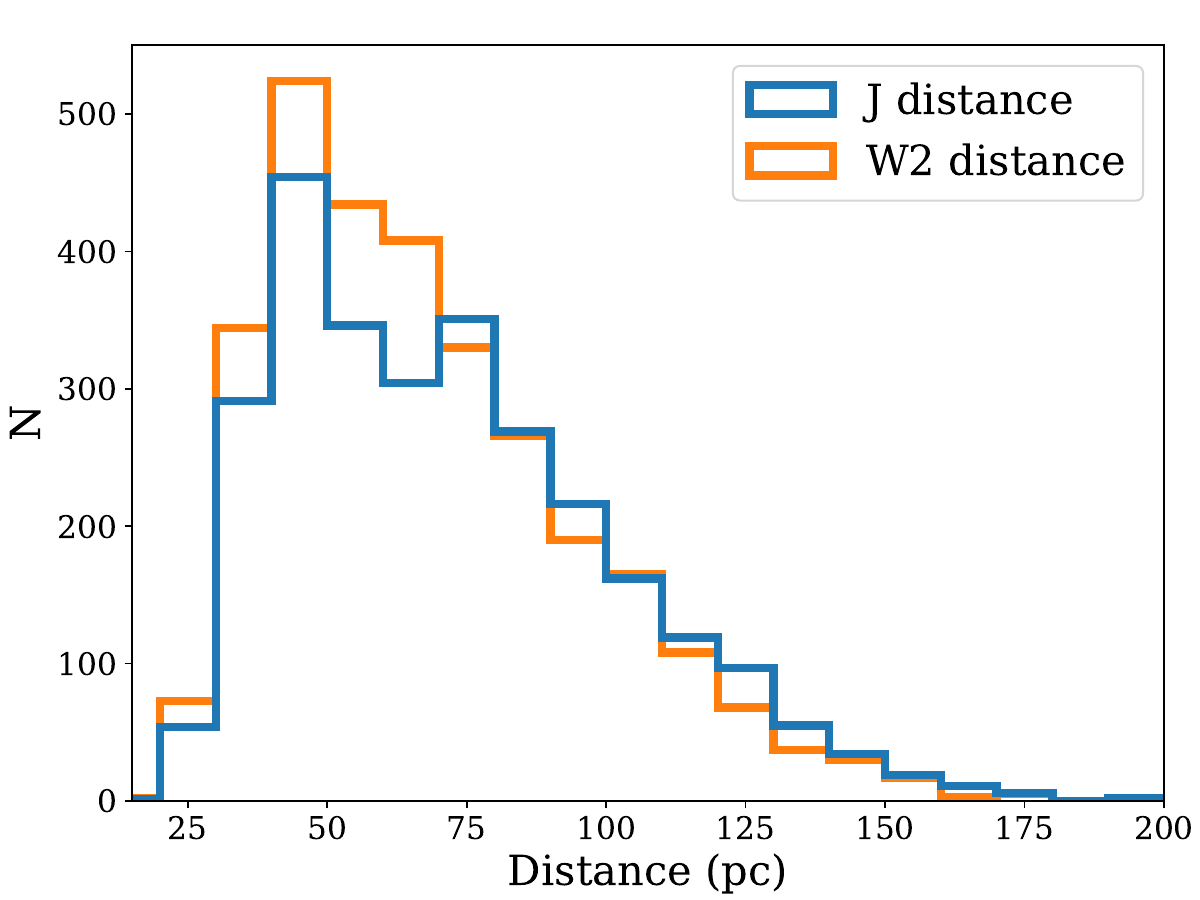}
\caption{Histogram of $J$-band (blue) and W2 (orange) distance estimates for the motion-confirmed BYW sample. }  
\label{fig:disthist}
\end{figure}

Distances are estimated for each object in the sample using relations from \cite{kirkpatrick2021} for available $J$-band and W2-band photometry. We use $J$ and W2 because these measurements are available for the majority of sources and generally bracket the red and blue edges of the available spectral energy distribution for each object.  Distance estimates are provided in Table \ref{tab:props} and their distribution is shown in Figure \ref{fig:disthist}.

Only 25 of our candidate ultracool dwarfs have distance estimates $\leq$25 pc. The lack of new nearby candidates is primarily because the vast majority of our nearest candidates have been published previously in \cite{kirkpatrick2021} and \cite{kirkpatrick2024}.  Three potentially nearby sources are CWISE J170226.68$-$224844.2, which is a WISE-only object with a W2 distance estimate of $\sim$13 pc based on a spectral type estimate of T8.5, CWISE J053238.12$-$013604.5, with a W2 distance estimate of 18 pc ($J$-band distance = 24 pc) based on an L7 spectral type estimate, and CWISE J011019.03+713348.5 with a $J$-band distance estimate of 19 pc (W2 distance = 27 pc) based on a T1 spectral type estimate.

\section{Notes on Particular Populations of Interest} 
\label{sec:objects}

Several subsets of this new catalog of discoveries will be investigated and analyzed in greater detail in future papers, including young objects and low-metallicity subdwarfs.  We do, however, highlight new potential multiple systems, including objects comoving with higher-mass primaries and low-mass binary pairs.  

\subsection{Wide Companions}

Wide L and T dwarf companions to main sequence stars and white dwarfs are often used as valuable benchmarks for constraining brown dwarf properties as well as substellar evolutionary models (e.g., \citealt{faherty2010, zhang2010, luhman2012, deacon2014, marocco2017, dalponte2020, zhang2020, bonavita2022}).  As such, wide companions have been the focus of several BYW papers, most notably \cite{rothermich2024}, who presented 89 such new systems, but also \cite{meisner2020, faherty2021, jalowiczor2021a, kiwy2021, schneider2021, gramaize2022, kiwy2022, marocco2024, bravo2025}.

During the visual inspection process for the candidates in this work, several sources were noted as having possible comoving companions.  The Gaia overlay option in WiseView is especially useful for identifying such systems with at least one Gaia detected source.  We have identified 28 systems with possible comoving, higher-mass primaries to new L and T dwarf candidates.  These 28 newly identified systems are listed in Table \ref{table:comovers1}.  Properties for the potential lower-mass companions are taken from Table \ref{tab:props}, while properties for the putative primaries, such as spectral type, proper motion, and distance, are taken from the literature as noted in Table \ref{table:comovers1}.  

Many of the primaries in these systems do not have directly measured spectral types.  We determined photometric types for all Gaia sources without existing spectral types by comparing their absolute G-band magnitudes from Gaia DR3 \citep{gaia2023} to the average G-band absolute magnitudes per spectral type found for nearby sources in \cite{kirkpatrick2024}.  These types are listed in square brackets in Table \ref{table:comovers1}.

All possible pairs were evaluated with the \texttt{CoMover} tool \citep{gagne2021} to determine their probabilities of being physically linked systems.  The \texttt{CoMover} code uses the available kinematics (e.g., proper motion, parallax, radial velocity) of two sources and determines a probability using a Bayesian framework that the two input sources are a physically associated pair by comparing the kinematics of the putative secondaries to those of the potential host stars and a field star model from \cite{gagne2018}. \texttt{CoMover} probabilities are listed in the last column of Table \ref{table:comovers1}. The \texttt{CoMover} percentage listed does not use a photometric distance for the secondary. 

\begin{figure*}
\epsscale{1.18}
\plotone{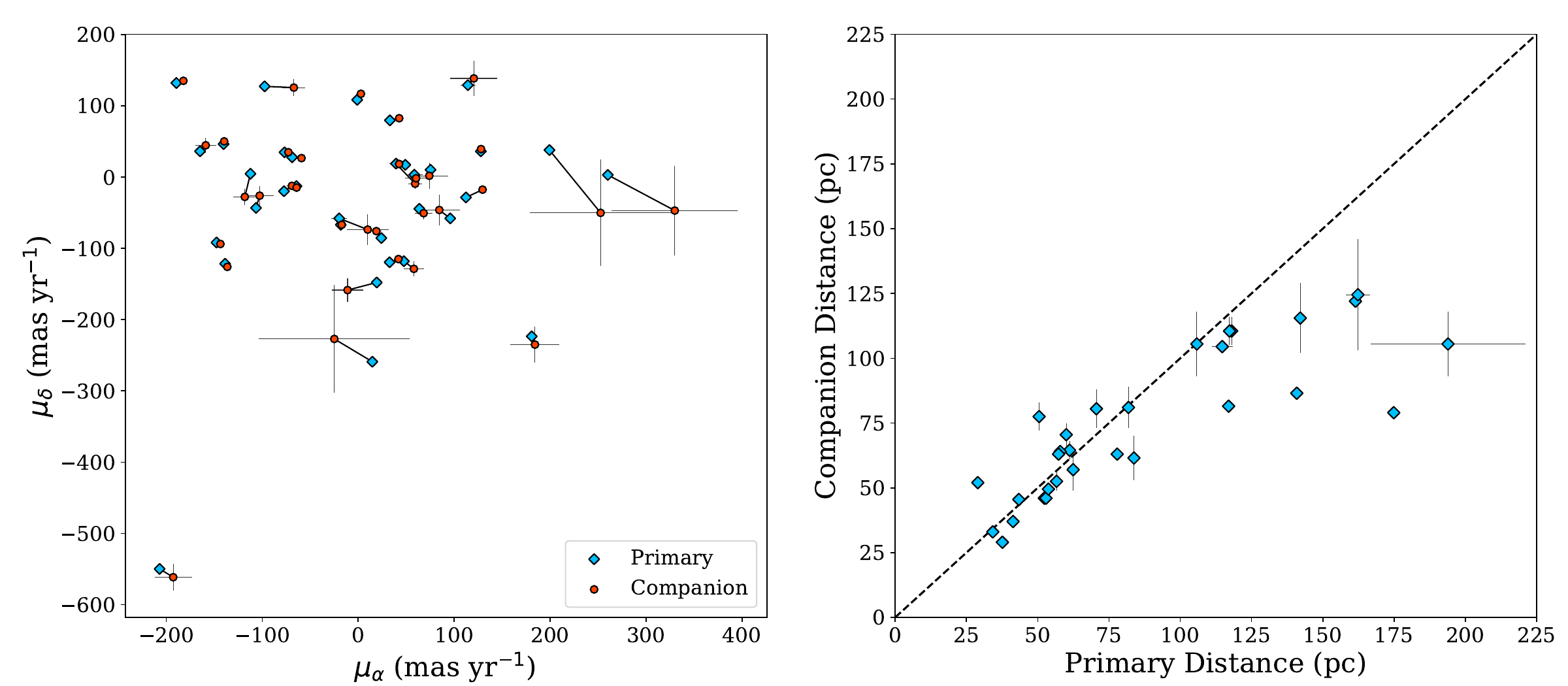}
\caption{Left: Proper motions components of newly identified systems, with primaries represented by blue diamonds and candidate companions plotted as red circles.  Lines connect the components of each system.  Right: Comparison of primary distance measurements versus photometric distance estimates of candidate companions, where the dashed black line represents equivalent distances.  Median values are plotted for photometric distances, with error bars representing the entire range of estimated distances.  Most distance estimates are consistent with measured primary distances, with some discrepancies for the most distant pairs, likely due to uncertain spectral type estimates and/or larger distance uncertainties.} 
\label{fig:companions}
\end{figure*}

Figure \ref{fig:companions} compares the proper motion components of these new potential pairs, highlighting their similar proper motion measurements.  For those few cases where the difference between proper motion measurements is larger, the uncertainties of the measurements are typically larger as well. Figure \ref{fig:companions} also shows a comparison of the measured distances of the primaries to the photometric distance estimates for candidate companions from Section \ref{sec:dist}.  While most distance estimates are consistent between primaries and companions, there are some significant discrepancies at larger distances ($\gtrsim$100 pc).  These could be due to inaccurate spectral type estimates, which would lead to incorrect distance estimates, or larger distance uncertainties at greater distances.  More accurate spectral type determinations and/or more precise distance measurements could help reconcile these inconsistencies.

We provide more details for some of the particularly interesting new systems in the following sections.

\subsubsection{LP 704$-$77 and CWISE J001822.14--135823.1} 
LP 704$-$77 has an exceptionally detailed inventory of chemical abundances from \cite{ting2019}, who determined a low-metallicty for this source ([Fe/H] = -0.715 dex).  This star has also been suggested as a member of one of the newly identified halo clusters in \cite{lovdal2022}.   

\subsubsection{CD--42 565AB and CWISE J013745.05--420956.8}
CD$-$42 565A and CD$-$42 565B were identified as a wide binary pair in \cite{oh2017} and \cite{andrews2017}, making this a potentially higher order system. \cite{bochanski2018} determined a log(age) for CD$-$42 565A of 9.78$^{+0.15}_{-0.11}$ Gyr and 9.85$^{+0.24}_{-0.20}$ Gyr for CD$-$42 565B using the \texttt{isochrones} package \citep{morton2015}. \cite{bochanski2018} further determined [Fe/H] values of 0.10$^{+0.14}_{-0.11}$ dex and 0.05$\pm$0.09 dex for A and B (respectively). 

\subsubsection{PM J02155--2437 and CWISE J021530.62--243612.1}
PM J02155$-$2437 is listed as a candidate binary in \cite{medan2023} based on Gaia and 2MASS photometry.  A second Gaia source (Gaia DR3 5119841470576786560) is located $\sim$1\arcsec\ from PM J02155$-$2437, but it does not have parallax or proper motion measurements.  Gaia DR3 gives a Renormalized Unit Weight Error (RUWE) value of 1.944, which is well over the 1.4 threshold suggested in \cite{lindegren2018} to separate potential multiple sources, which supports the binary hypothesis for the primary in this system. 

\subsubsection{UCAC4 297--003544 and CWISE J033205.98--304234.6}
UCAC4 297$-$003544 was suggested as an AB Dor candidate in \cite{gagne2015} based on the proper motion of this source at that time ($\sim$50\% probability of membership).  While this object does have a position and photometry in Gaia DR3, there is no Gaia parallax or proper motion available for this source.  Using the radial velocity from RAVE DR6 (24.09$\pm$4.81 \kms) and an updated proper motion from UCAC5 \citep{zacharias2017}, the probability of AB Dor membership increases to 88.4\% using the most recent BANYAN version.  We also note that the $J$-band and W2 photometric distances for the possible L6.5 companion, CWISE J033205.98$-$304234.6, are highly discrepant (d$_{\rm J}$=120.9 pc, d$_{\rm W2}$=53.2 pc), a trait common to very red L dwarfs. This is supported by this object's exceptionally red colors ($J-K$ = 2.48$\pm$0.10 mag, $J-$W2 = 4.53$\pm$0.08 mag). Note also that young L dwarfs tend to have redder colors than their field counterparts of the same type, so the red colors of this source are consistent with this system potentially being a member of the AB Dor moving group. We therefore determine an approximate distance to this system using the $K$-band photometric distance of CWISE J033205.98$-$304234.6 (81.7 pc), which has been shown to be most accurate for very red L dwarfs with known distances \citep{schneider2023}.  This distance is used to determine the separation of this system in au and for input into \texttt{CoMover}. 

\subsubsection{G 160--45 and CWISE J040633.62--170849.6}
G 160$-$45 has a Gaia DR3 RUWE value of 1.449, just over 1.4 threshold suggested to be indicative of binarity, making this a potential higher-order multiplicity system.  Note, however, that the distribution of RUWE values for M type stars is centered around a higher value ($\sim$1.2) than those of stars with bluer colors \citep{sozzetti2023}.  We estimate a spectral type of $\sim$M1 for G 160--45, and thus the evidence for multiplicity for this system may have less significance. 

\subsubsection{StKM 1--570AB and CWISE J053409.85+651953.1}
StKM 1$-$570 was noted as a ROSAT x-ray source \citep{haakonsen2009}, and later found to be a field binary in \cite{bowler2019}. Gaia DR3 RUWE values are 1.660 and 1.529 for the A and B components, respectively, consistent with the multiplicity of this system.   

\subsubsection{ATO J119.4548+55.4544, Gaia DR3 1080989465348498816, and CWISE J075748.41+552712.1}
CWISE J075748.41+552712.1 was noted to have a very similar proper motion to two sources based on their Gaia astrometry,  ATO J119.4548+55.4544 and Gaia DR3 1080989465348498816.  However, the parallax measurements for these two sources are inconsistent with being a physically associated system.  We note however, that ATO J119.4548+55.4544 and Gaia DR3 1080989465348498816 were included as a possible wide binary system in \cite{hartman2020}, and we therefore list both objects as possible companions to CWISE J075748.41+552712.1.

\subsubsection{12 Sex and CWISE J095945.81+032530.0}
12 Sex has been considered an astrometric binary candidate since \cite{makarov2005}, and was found to have a proper motion acceleration in \cite{brandt2021}.  The Gaia DR3 RUWE value of 3.110 supports the likely multiplicity of this source, though the wide separation and likely low-mass of the possible secondary CWISE J095945.81+032530.0 would not account for these astrometric signs of binarity.  12 Sex also has an age determination of 641$^{+688}_{-262}$ Myr from \cite{david2015}.  

\subsubsection{UCAC4 683--055294 and CWISE J122353.94+462616.7}
UCAC4 683$-$055294 has been suggested as a possible member of Crius 222, which has an age of 0.1-0.7 Gyr \citep{moranta2022}. UCAC4 683$-$055294 also has a Gaia DR3 RUWE value of 1.969, making this system a candidate for higher-order multiplicity.   

\subsubsection{Gaia DR3 1282869627093730304 and CWISE J144531.27+304107.3}
Gaia DR3 1282869627093730304 and CWISE J144531.27+304107.3 are listed as a possible wide binary in \cite{hartman2020}.

\subsubsection{G 144--61AB and CWISE J210510.74+193510.4}
G 144$-$61 is listed as a visual double in \cite{dommanget2000} using Hipparcos data, and is later confirmed as a close binary in \cite{horch2010} with a separation of 0\farcs347.

\startlongtable
\begin{longrotatetable}
\begin{deluxetable*}{lcccccccccccccc}
\tabletypesize{\scriptsize}
\tablecaption{Potential New Comoving Systems}
\label{table:comovers1}
\tablehead{
\colhead{Name} & 
\colhead{$\alpha$} & 
\colhead{$\delta$} & 
\colhead{Pos.} &
\colhead{SpType\tablenotemark{a}} &
\colhead{SpType} &
\colhead{$\mu_{\alpha}$} & 
\colhead{$\mu_{\delta}$} & 
\colhead{Dist.\tablenotemark{b}} &
\colhead{Ast.} &
\colhead{Sep} &
\colhead{Sep\tablenotemark{c}} &
\colhead{\texttt{CoMover}} \\
 & \colhead{($\degr$)} & \colhead{($\degr$)} & \colhead{Ref.} & & \colhead{Ref.} & \colhead{(\mas)} & \colhead{(\mas)} & \colhead{(pc)} & \colhead{Ref.} & \colhead{($\arcsec$)} & \colhead{(au)} & \colhead{($\%$)}}
\startdata
LP 704-77	&	4.5848726	&	-13.9740472	&	1	&	[M4]	&	3	&	199.300	$\pm$	0.037	&	38.013	$\pm$	0.026	&	56.61	$\pm$	0.13	&	1	&	\dots	&	\dots	&	95.8	\\	
 CWISE J001822.14-135823.1	&	4.5922541	&	-13.9730982	&	2	&	[T4]	&	4	&	252.6	$\pm$	73.8	&	-50.1	$\pm$	74.6	&	[49--56]			&	2	&	26	&	1470	&	\dots	\\	\hline
[PS78] 9	&	11.3503556	&	-24.2880358	&	1	&	[M5]	&	3	&	33.023	$\pm$	0.042	&	79.711	$\pm$	0.031	&	34.24	$\pm$	0.04	&	1	&	\dots	&	\dots	&	99.9	\\	
CWISE J004522.21-241650.3	&	11.3425570	&	-24.2806502	&	2	&	[T3.5]	&	4	&	42.7	$\pm$	3.1	&	82.7	$\pm$	3.2	&	[31--35]			&	10	&	37	&	1260	&	\dots	\\	\hline
Gaia DR3 2787961957297168256	&	14.2168553	&	18.2041199	&	1	&	[M2]	&	3	&	127.765	$\pm$	0.025	&	36.264	$\pm$	0.023	&	60.01	$\pm$	0.08	&	1	&	\dots	&	\dots	&	100.0	\\	
CWISE J005652.93+181157.9	&	14.2205566	&	18.1994154	&	2	&	[L2]	&	4	&	128.0	$\pm$	2.4	&	39.4	$\pm$	2.4	&	[66--75]			&	11	&	21	&	1270	&	\dots	\\	\hline
CD-42 565A	&	24.4100092	&	-42.1831184	&	1	&	K0	&	5	&	112.287	$\pm$	0.011	&	-28.488	$\pm$	0.013	&	118.04	$\pm$	0.22	&	1	&	\dots	&	\dots	&	4.2	\\	
CD-42 565B	&	24.4068280	&	-42.1865027	&	1	&	[K2.5]	&	3	&	113.905	$\pm$	0.010	&	-28.235	$\pm$	0.013	&	117.28	$\pm$	0.21	&	1	&	15	&	1750	&	26.2	\\	
CWISE J013745.05-420956.8	&	24.4377395	&	-42.1657986	&	2	&	[L2]	&	4	&	129.6	$\pm$	2.9	&	-17.5	$\pm$	3.1	&	[105--116]			&	10	&	97	&	11420	&	\dots	\\	\hline
PM J02155-2437	&	33.8991014	&	-24.6221132	&	1	&	[M1.5]	&	3	&	32.691	$\pm$	0.084	&	-119.444	$\pm$	0.091	&	61.17	$\pm$	0.40	&	1	&	\dots	&	\dots	&	99.8	\\	
CWISE J021530.62-243612.1	&	33.8776285	&	-24.6033740	&	2	&	[L1.5]	&	4	&	41.9	$\pm$	3.0	&	-115.0	$\pm$	3.1	&	[61--68]			&	10	&	97	&	5960	&	\dots	\\	\hline
LP 355-25	&	46.7376701	&	22.6879213	&	1	&	M2	&	6	&	14.718	$\pm$	0.017	&	-259.057	$\pm$	0.013	&	37.61	$\pm$	0.02	&	1	&	\dots	&	\dots	&	55.2	\\	
CWISE J030651.26+224000.8	&	46.7136218	&	22.6669087	&	2	&	[T7.5]	&	4	&	-24.9	$\pm$	78.5	&	-227.1	$\pm$	75.7	&	[28--30]			&	14	&	110	&	4140	&	\dots	\\	\hline
UCAC4 297-003544\tablenotemark{d}	&	53.0379646	&	-30.7116268	&	1	&	[M1.5]	&	3	&	63.7	$\pm$	1.7	&	-44.4	$\pm$	1.7	&	\dots			&	15	&	\dots	&	\dots	&	99.7	\\	
CWISE J033205.98-304234.6	&	53.0249675	&	-30.7096211	&	2	&	[L6.5]	&	4	&	68.2	$\pm$	9.0	&	-50.6	$\pm$	9.0	&	[53--121]			&	14	&	41	&	3352	&	\dots	\\	\hline
G 160-45	&	61.6349444	&	-17.1447254	&	1	&	[M1]	&	3	&	180.861	$\pm$	0.019	&	-223.457	$\pm$	0.019	&	41.34	$\pm$	0.03	&	1	&	\dots	&	\dots	&	99.9	\\	
CWISE J040633.62-170849.6	&	61.6400934	&	-17.1471160	&	2	&	[T5]	&	4	&	184.2	$\pm$	25.6	&	-235.0	$\pm$	25.5	&	[36--38]			&	10	&	20	&	810	&	\dots	\\	\hline
Gaia DR3 3296800270594251392	&	62.4007688	&	5.7268759	&	1	&	[K5]	&	3	&	49.110	$\pm$	0.019	&	17.310	$\pm$	0.012	&	161.45	$\pm$	0.44	&	1	&	\dots	&	\dots	&	99.6	\\	
CWISE J040935.09+054344.7	&	62.3962138	&	5.7290864	&	2	&	[L1]	&	4	&	42.6	$\pm$	2.9	&	18.5	$\pm$	2.9	&	[121--123]			&	11	&	18	&	2930	&	\dots	\\	\hline
Gaia DR3 4899230990814961408	&	65.9806469	&	-21.7994850	&	1	&	[M0]	&	3	&	19.325	$\pm$	0.012	&	-147.814	$\pm$	0.014	&	77.88	$\pm$	0.09	&	1	&	\dots	&	\dots	&	99.6	\\	
CWISE J042358.54-214839.8	&	65.9939464	&	-21.8110931	&	2	&	[T4.5]	&	4	&	-11.1	$\pm$	16.6	&	-158.5	$\pm$	16.6	&	[62--64]			&	10	&	61	&	4750	&	\dots	\\	\hline
StKM 1-570A	&	83.7457314	&	65.3615497	&	1	&	M0	&	7	&	47.591	$\pm$	0.026	&	-117.962	$\pm$	0.030	&	52.32	$\pm$	0.10	&	1	&	\dots	&	\dots	&	98.4	\\	
StKM 1-570B	&	83.7449549	&	65.3615810	&	1	&	[M0]	&	3	&	51.541	$\pm$	0.030	&	-121.259	$\pm$	0.040	&	52.89	$\pm$	0.13	&	1	&	1.2	&	60	&	99.0	\\	
CWISE J053409.85+651953.1	&	83.5410791	&	65.3314362	&	2	&	[L8.5]	&	4	&	58.0	$\pm$	10.8	&	-128.7	$\pm$	10.4	&	[46]			&	14	&	326	&	17050	&	\dots	\\	\hline
ATO J119.4548+55.4544	&	119.4547893	&	55.4544319	&	1	&	M5	&	6	&	-18.171	$\pm$	0.038	&	-66.840	$\pm$	0.040	&	105.82	$\pm$	0.52	&	1	&	\dots	&	\dots	&	99.6	\\	
CWISE J075748.41+552712.1	&	119.4517392	&	55.4533694	&	2	&	[L3]	&	4	&	-17.2	$\pm$	4.0	&	-66.8	$\pm$	4.0	&	[93--118]			&	11	&	7.3	&	770	&	\dots	\\	
Gaia DR3 1080989465348498816	&	119.4270354	&	55.4533428	&	1	&	[M7.5]	&	3	&	-18.791	$\pm$	0.697	&	-69.558	$\pm$	0.636	&	193.91	$\pm$	27.15	&	1	&	57	&	6010	&	99.8	\\	\hline
SCR J0823-6957	&	125.7562893	&	-69.9500359	&	1	&	[M3.5]	&	3	&	-189.495	$\pm$	0.017	&	132.157	$\pm$	0.017	&	43.38	$\pm$	0.02	&	1	&	\dots	&	\dots	&	100.0	\\	
CWISE J082302.50-695715.0	&	125.7604486	&	-69.9541622	&	2	&	[L7]	&	4	&	-182.2	$\pm$	4.1	&	135.2	$\pm$	4.2	&	[44--47]			&	14	&	16	&	680	&	\dots	\\	\hline
Gaia DR3 1089909665747345920	&	126.4506186	&	60.9783482	&	1	&	[M2]	&	3	&	75.364	$\pm$	0.011	&	10.394	$\pm$	0.013	&	57.89	$\pm$	0.05	&	1	&	\dots	&	\dots	&	93.9	\\	
CWISE J082551.72+605727.8	&	126.4654869	&	60.9577299	&	2	&	[L8]	&	4	&	74.1	$\pm$	19.7	&	1.9	$\pm$	18.4	&	[64]			&	14	&	79	&	4550	&	\dots	\\	\hline
Gaia DR3 616825649520497280	&	148.3923996	&	15.9846872	&	1	&	M4	&	6	&	-77.323	$\pm$	0.034	&	-19.751	$\pm$	0.028	&	70.59	$\pm$	0.17	&	1	&	\dots	&	\dots	&	98.1	\\	
CWISE J095333.72+155933.0	&	148.3905097	&	15.9925232	&	2	&	[L6.5]	&	4	&	-69.1	$\pm$	5.5	&	-12.2	$\pm$	5.5	&	[73--88]			&	11	&	29	&	2040	&	\dots	\\	\hline
12 Sex	&	149.9292116	&	3.3849156	&	1	&	A5	&	8	&	-68.845	$\pm$	0.127	&	27.905	$\pm$	0.133	&	81.83	$\pm$	0.64	&	1	&	\dots	&	\dots	&	96.7	\\	
CWISE J095945.81+032530.0	&	149.9409098	&	3.4250033	&	2	&	[L4]	&	4	&	-59.1	$\pm$	3.2	&	26.7	$\pm$	3.2	&	[73--89]			&	12	&	150	&	12300	&	\dots	\\	\hline
Gaia DR3 3485780991984532352	&	174.3907865	&	-24.7814752	&	1	&	[M4.5]	&	3	&	24.078	$\pm$	0.104	&	-85.432	$\pm$	0.085	&	140.87	$\pm$	2.18	&	1	&	\dots	&	\dots	&	99.8	\\	
CWISE J113734.09-244703.3	&	174.3920409	&	-24.7842583	&	2	&	[L2]	&	4	&	18.9	$\pm$	5.3	&	-75.8	$\pm$	5.3	&	[84--89]			&	16	&	11	&	1530	&	\dots	\\	\hline
UCAC4 683-055294	&	186.0035575	&	46.4395070	&	1	&	K7	&	9	&	-64.310	$\pm$	0.014	&	-12.596	$\pm$	0.016	&	83.75	$\pm$	0.15	&	1	&	\dots	&	\dots	&	99.7	\\	
CWISE J122353.94+462616.7	&	185.9747487	&	46.4379827	&	2	&	[L4]	&	4	&	-64.2	$\pm$	2.1	&	-14.6	$\pm$	2.1	&	[53--70]			&	11	&	72	&	6000	&	\dots	\\	\hline
Gaia DR3 6186489795309499392	&	198.5582369	&	-27.2446449	&	1	&	[M4.5]	&	3	&	-106.470	$\pm$	0.166	&	-43.207	$\pm$	0.139	&	162.31	$\pm$	4.16	&	1	&	\dots	&	\dots	&	99.7	\\	
CWISE J131414.03-271446.5	&	198.5584597	&	-27.2462789	&	2	&	[L1]	&	4	&	-102.7	$\pm$	14.5	&	-26.0	$\pm$	13.4	&	[103--146]			&	14	&	5.9	&	960	&	\dots	\\	\hline
LSPM J1344+3753	&	206.2063494	&	37.8854045	&	1	&	M2	&	6	&	-164.841	$\pm$	0.019	&	36.286	$\pm$	0.024	&	53.75	$\pm$	0.09	&	1	&	\dots	&	\dots	&	99.9	\\	
CWISE J134451.12+375307.6	&	206.2130284	&	37.8854542	&	2	&	[T5]	&	4	&	-158.9	$\pm$	10.8	&	44.7	$\pm$	10.8	&	[47--52]			&	11	&	19	&	1020	&	\dots	\\	\hline
Gaia DR3 6275160803885682688	&	209.1792245	&	-23.8723024	&	1	&	[M4.5]	&	3	&	-112.299	$\pm$	0.274	&	4.770	$\pm$	0.319	&	62.42	$\pm$	0.92	&	1	&	\dots	&	\dots	&	89.8	\\	
CWISE J135642.15-235216.5	&	209.1756314	&	-23.8712844	&	2	&	[L7]	&	4	&	-118.2	$\pm$	11.8	&	-27.7	$\pm$	11.3	&	[49--65]			&	14	&	12	&	770	&	\dots	\\	\hline
Gaia DR3 1282869627093730304	&	221.3781418	&	30.6853321	&	1	&	M4	&	6	&	-140.298	$\pm$	0.038	&	46.418	$\pm$	0.042	&	174.89	$\pm$	1.25	&	1	&	\dots	&	\dots	&	100.0	\\	
CWISE J144531.27+304107.3	&	221.3804926	&	30.6851311	&	1	&	[L2]	&	4	&	-139.7	$\pm$	1.0	&	50.2	$\pm$	1.0	&	[77--81]			&	1	&	7.3	&	1280	&	\dots	\\	\hline
Gaia DR3 1273955714528239488	&	232.0353923	&	31.3757351	&	1	&	[M6.5]	&	3	&	-76.776	$\pm$	0.254	&	34.862	$\pm$	0.305	&	114.71	$\pm$	3.63	&	1	&	\dots	&	\dots	&	100.0	\\	
CWISE J152807.08+312202.6	&	232.0295261	&	31.3674047	&	2	&	[L4.5]	&	4	&	-72.8	$\pm$	2.2	&	35.0	$\pm$	2.2	&	[103--106]			&	11	&	35	&	4010	&	\dots	\\	\hline
Gaia DR3 4402371100793892736	&	235.7778690	&	-4.0252308	&	1	&	[K6]	&	3	&	-147.669	$\pm$	0.017	&	-91.707	$\pm$	0.014	&	116.96	$\pm$	0.22	&	1	&	\dots	&	\dots	&	100.0	\\	
CWISE J154305.45-040126.5	&	235.7727415	&	-4.0240505	&	2	&	[L0.5]	&	4	&	-143.6	$\pm$	2.8	&	-93.8	$\pm$	2.8	&	[79--84]			&	16	&	19	&	2210	&	\dots	\\	\hline
LSPM J1630+8252	&	247.7310846	&	82.8757047	&	1	&	[M5.5]	&	3	&	-97.499	$\pm$	0.086	&	127.440	$\pm$	0.083	&	57.30	$\pm$	0.21	&	1	&	\dots	&	\dots	&	94.2	\\	
CWISE J163129.24+824746.7	&	247.8718570	&	82.7963193	&	2	&	[L9]	&	4	&	-67.1	$\pm$	12.0	&	125.7	$\pm$	12.3	&	[63]			&	14	&	293	&	16770	&	\dots	\\	\hline
UCAC4 180-220563	&	291.4276902	&	-54.0074966	&	1	&	[M3.5]	&	3	&	260.152	$\pm$	0.104	&	3.101	$\pm$	0.099	&	28.98	$\pm$	0.12	&	1	&	\dots	&	\dots	&	99.8	\\	
CWISE J192539.85-540036.6	&	291.4160609	&	-54.0101867	&	2	&	[T5.5]	&	4	&	329.9	$\pm$	65.6	&	-46.9	$\pm$	62.6	&	[52]			&	14	&	26	&	770	&	\dots	\\	\hline
Gaia DR3 2047812098630732160	&	295.8986579	&	35.6243514	&	1	&	[M4]	&	3	&	-1.052	$\pm$	0.055	&	108.605	$\pm$	0.060	&	142.09	$\pm$	1.09	&	1	&	\dots	&	\dots	&	100.0	\\	
CWISE J194335.80+353722.8	&	295.8991800	&	35.6230100	&	2	&	[L3.5]	&	4	&	2.8	$\pm$	2.6	&	117.0	$\pm$	2.6	&	[102--129]			&	11	&	5.1	&	720	&	\dots	\\	\hline
G 144-61AB	&	316.2876798	&	19.5992496	&	1	&	[M0]	&	3	&	-206.947	$\pm$	0.193	&	-549.866	$\pm$	0.163	&	50.48	$\pm$	1.50	&	18	&	\dots	&	\dots	&	100.0	\\	
CWISE J210510.74+193510.4	&	316.2947711	&	19.5862389	&	2	&	[T4]	&	4	&	-192.8	$\pm$	19.4	&	-561.5	$\pm$	19.2	&	[72--83]			&	19	&	53	&	2660	&	\dots	\\					
\enddata
\tablerefs{(1) Gaia DR3 \citep{gaia2023}; (2) CatWISE2020 \citep{marocco2021}; (3) this work and \cite{kirkpatrick2024}; (4) this work; (5) \cite{SpencerJones1939}; (6) \citep{li2024}; (7) \cite{klutsch2020}; (8) \cite{morgan1952}; (9) \cite{zhang2023}; (10) NSC DR2 \citep{nidever2021}; (11) UHS DR3 \citep{schneider2025}; (12) UKIDSS \citep{lawrence2007}; (13) VIKING detections \citep{edge2013} (this work) (14) unTimely detections \citep{meisner2023} (this work); (15) UCAC5 \citep{zacharias2017}; (16) PS1 detections \citep{chambers2016} (this work); (17) VHS detections \citep{mcmahon2013} (this work); (18) Gaia DR1 \citep{gaia2016}; (19) UHS detections \citep{dye2018} (this work)  }
\tablenotetext{a}{Spectral types in square brackets are photometric estimates.}
\tablenotetext{b}{Distances in square brackets are photometric distance estimates as described in Section \ref{sec:dist}.}
\tablenotetext{c}{Separations in au are based on the distances to the primaries in each system.}
\tablenotetext{d}{No parallax is available for the UCAC4 297$-$003544 and CWISE J033205.98$-$304234.6 system, so we use the $K$-band photometric distance of CWISE J033205.98$-$304234.6 to determine the separation of the system in au and in the \texttt{CoMover} evaluation (see text). }
\end{deluxetable*} 
\end{longrotatetable}

\subsection{Close Cold Companions}
Binaries consisting of resolved L, T, or Y type components are rare (e.g., \citealt{burgasser2006, reid2008, pope2013, aberasturi2014, fontanive2018, fontanive2023, factor2023}), with discoveries of such ultracool binaries typically coming from high-resolution imaging with the Hubble Space Telescope (e.g., \citealt{reid2001, reid2006, bouy2003, burgasser2003, gizis2003, ryan2011, factor2022, mehta2026}), adaptive optics from the ground (e.g., \citealt{close2002, mccoughrean2004, liu2005, liu2006, liu2010, liu2011, liu2012, goldman2008, looper2008, allers2010, bernat2010, stumpf2010, gelino2011, gelino2014, artigau2011, dupuy2012, radigan2013, dupuy2015, melso2015, bardalez2015, opitz2016, best2017, baig2025, xuan2024}), or more recently imaging with the James Webb Space Telescope (e.g., \citealt{calissendorff2023, defurio2025}).

Only a small number of binaries with L and T components have been found that have on-sky separations wide enough to be detected in large-area survey images, including SDSS J14162408+1348263AB (L7+T7.5; \citealt{bowler2010, burningham2010, schmidt2010}), WISE 2150$-$7520AB (L1+T8; \citealt{faherty2020}), 2MASS J0139+8110AB (L1+L2; \citealt{marocco2020}), and CWISE J014611.20$-$050850.0AB (L4+L8; \citealt{softich2022}). 

\begin{figure*}
\epsscale{1.18}
\plotone{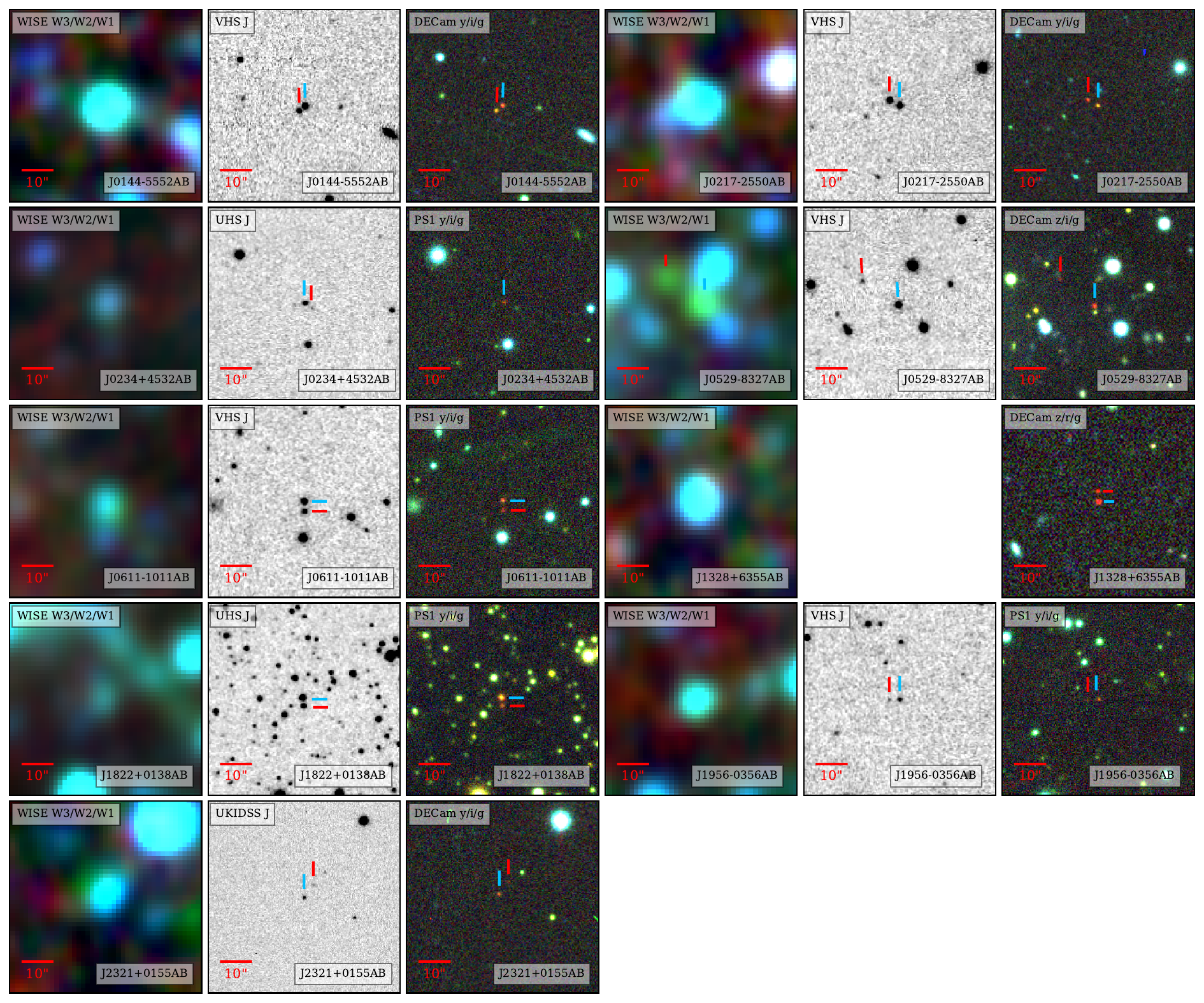}
\caption{WISE, near-infrared, and optical images of all ultracool wide binary candidates. Optical images either come from Pan-STARRS (PS1; \citealt{chambers2016}) or DECam images obtained through the Astro Data Lab image cutout service \citep{fitzpatrick2014, nikutta2020}. Primaries are marked with light blue lines while candidate ultracool companions are marked with red lines in all images where the components are resolved. Neither component of the CWISE J132836.50+635527.7AB system is detected in any available near-infrared imaging.  North is up and East is left in all images.} 
\label{fig:cutouts1}
\end{figure*}

We looked for close binaries similar to the SDSS J14162408+1348263AB, WISE 2150$-$7520AB, 2MASS J0139+8110AB, and CWISE J014611.20$-$050850.0AB systems using available optical and near-infrared imaging for each source in our sample.  We discovered nine candidate systems, six of which are supported via similar proper motion measurements.  The six systems with consistent proper motion measurements are plotted in the left panel of Figure \ref{fig:companions}.  Information for these potential new systems is provided in Table \ref{table:comovers2}.  Cutouts showing the positions of the components in these systems are shown in Figure \ref{fig:cutouts1}.

For one of the remaining systems not supported via similar proper motions (CWISE J052901.09$-$832749.1 and CWISE J052907.78$-$832740.9), the proper motion of the putative secondary (CWISE J052907.78$-$832740.9), while consistent with the primary to within 1$\sigma$ (combined), is itself not significant (e.g., $<$3$\sigma$).  The photometric distances determined for these two sources in Section \ref{sec:dist} are consistent, however, considering the uncertainties in their spectral type estimates (34 to 36 pc for CWISE J052901.09$-$832749.1 and 29 to 31 pc for CWISE J052907.78$-$832740.9).  

For the remaining two systems (CWISE J132836.50+635527.7AB and CWISE J195612.41-035615.7AB), proper motions were not able to be measured for the putative secondaries. 

Spectral type estimates for secondaries are determined using $J$- and $K$-band photometry combined with $J$-band distance estimates of the primaries compared to absolute $J$- and $K$-band magnitudes as a function of spectral type from \cite{kirkpatrick2021}.  For CWISE J132836.50+635527.7B, no near-infrared photometry exists, so we used a $z$-band measurement from NSC DR2 \citep{nidever2021} of 21.409$\pm$0.077 mag, the $J$-band distance estimate for CWISE J132836.50+635527.7A, and a comparison to Table 6 of \cite{best2018} to determine a spectral type estimate of T2.

We ran \texttt{CoMover} for the potential systems in Table \ref{table:comovers2}, using the $J$-band photometric distances of the primaries but without using a photometric distance for the secondaries.  We assume a 20\% uncertainty for the photometric distances used, which should encapsulate the majority of the combined uncertainty for these sources from spectral type estimates, photometric uncertainties, and uncertainties from the spectral type versus absolute $J$-band magnitude relation.  Caution is needed when interpreting the resulting \texttt{CoMover} percentages considering the large uncertainties for the measured proper motions for most components as well as those associated with the distance estimates to each system.  We do not run \texttt{CoMover} on the two systems for which we could not measure a proper motion for the possible secondaries. 

Of particular note among this sample of new potential ultracool binaries is the pair CWISE J052901.09$-$832749.1 and CWISE J052907.78$-$832740.9.  While several binaries consisting of multiple T dwarfs have been found previously \citep{burgasser2003, burgasser2006, mccoughrean2004, liu2006, liu2010, liu2012, looper2008, artigau2011, gelino2011, radigan2013}, their projected physical separations are all $\lesssim$15 au, compared to $\sim$510 au for CWISE J052901.09$-$832749.1 and CWISE J052907.78$-$832740.9. More precise proper motion measurements and spectroscopic observations are needed to confirm these two objects as a physical pair.

\startlongtable
\begin{longrotatetable}
\begin{deluxetable*}{lcccccccccccccc}
\tabletypesize{\small}
\tablecaption{Ultracool Wide Binary Candidates}
\label{table:comovers2}
\tablehead{
\colhead{Name} & 
\colhead{$\alpha$} & 
\colhead{$\delta$} & 
\colhead{Pos.} &
\colhead{SpType\tablenotemark{a}} &
\colhead{SpType} &
\colhead{$\mu_{\alpha}$} & 
\colhead{$\mu_{\delta}$} & 
\colhead{Ast.} &
\colhead{Sep} &
\colhead{Sep\tablenotemark{b}} &
\colhead{\texttt{CoMover}} \\
 & \colhead{($\degr$)} & \colhead{($\degr$)} & \colhead{Ref.} & & \colhead{Ref.} & \colhead{(\mas)} & \colhead{(\mas)}  & \colhead{Ref.} & \colhead{($\arcsec$)} & \colhead{(au)} & \colhead{($\%$)}}
\startdata
CWISE J014425.00$-$555225.6A	&	26.1041171	&	-55.8737391	&	1	&	[L9.5]	&	6	&	39.4	$\pm$	6.7	&	19.0	$\pm$	6.7	&	1	&	\dots	&	\dots	&	30.4	\\	
CWISE J014425.00$-$555225.6B	&	26.1050181	&	-55.8741681	&	1	&	[T6]	&	6	&	59.3	$\pm$	7.1	&	-9.2	$\pm$	7.0	&	1	&	2.4	&	80	&	\dots	\\	\hline
CWISE J021729.71$-$255057.2A	&	34.3738008	&	-25.8492583	&	1	&	[L8.5]	&	6	&	58.5	$\pm$	8.7	&	3.3	$\pm$	8.7	&	1	&	\dots	&	\dots	&	73.3	\\	
CWISE J021729.71$-$255057.2B	&	34.3747655	&	-25.8487845	&	1	&	[T2]	&	6	&	60.3	$\pm$	11.8	&	-1.6	$\pm$	11.7	&	1	&	3.6	&	150	&	\dots	\\	\hline
CWISE J023406.08+453220.6A	&	38.5253658	&	45.5390564	&	2	&	[L7]	&	6	&	95.9	$\pm$	4.7	&	-57.9	$\pm$	4.7	&	2	&	\dots	&	\dots	&	82.8	\\	
CWISE J023406.08+453220.6B	&	38.5244974	&	45.5386214	&	2	&	[T7]	&	6	&	84.5	$\pm$	21.6	&	-46.1	$\pm$	21.6	&	2	&	2.7	&	140	&	\dots	\\	\hline
CWISE J052901.09$-$832749.1	&	82.2545656	&	-83.4636596	&	3	&	[T5.5]	&	6	&	148.6	$\pm$	24.2	&	-23.3	$\pm$	22.8	&	9	&	\dots	&	\dots	&	0.7	\\	
CWISE J052907.78$-$832740.9	&	82.2824140	&	-83.4613877	&	3	&	[T8]	&	6	&	76.4	$\pm$	48.3	&	-41.8	$\pm$	53.3	&	3	&	14.0	&	510	&	\dots	\\	\hline
CWISE J061133.39$-$101140.8A	&	92.8891565	&	-10.1944092	&	4	&	[L7.5]	&	6	&	-19.8	$\pm$	8.2	&	-57.9	$\pm$	7.8	&	8	&	\dots	&	\dots	&	13.9	\\	
CWISE J061133.39$-$101140.8B	&	92.8891366	&	-10.1952639	&	4	&	[T2]	&	6	&	9.9	$\pm$	21.7	&	-73.6	$\pm$	21.7	&	9	&	3.1	&	130	&	\dots	\\	\hline
CWISE J132836.50+635527.7A	&	202.1521018	&	63.9243803	&	3	&	[L4.5]	&	6	&	-103.4	$\pm$	6.5	&	1.4	$\pm$	6.3	&	8	&	\dots	&	\dots	&	\dots	\\	
CWISE J132836.50+635527.7B	&	202.1521814	&	63.9252544	&	1	&	[T2]	&	7	&	\dots			&	\dots			&	\dots	&	3.3	&	230\tablenotemark{c}	&	\dots	\\	\hline
CWISE J182241.97+013819.2A	&	275.6748056	&	1.6388732	&	2	&	[L3]	&	6	&	-138.7	$\pm$	2.1	&	-121.5	$\pm$	2.1	&	2	&	\dots	&	\dots	&	100	\\	
CWISE J182241.97+013819.2B	&	275.6747296	&	1.6381673	&	2	&	[L5]	&	6	&	-136.4	$\pm$	2.4	&	-125.9	$\pm$	2.4	&	2	&	2.6	&	140	&	\dots	\\	\hline
CWISE J195612.41$-$035615.7A	&	299.0508928	&	-3.9376068	&	4	&	[L6]	&	6	&	-51.3	$\pm$	7.5	&	-19.6	$\pm$	7.4	&	8	&	\dots	&	\dots	&	\dots	\\	
CWISE J195612.41$-$035615.7B	&	299.0518037	&	-3.9376809	&	4	&	[T1]	&	6	&	\dots			&	\dots			&	\dots	&	3.3	&	220	&	\dots	\\	\hline
CWISE J232159.46+015529.7A	&	350.4976039	&	1.9246156	&	5	&	[L5]	&	6	&	114.3	$\pm$	7.8	&	129.1	$\pm$	8.0	&	10	&	\dots	&	\dots	&	100	\\	
CWISE J232159.46+015529.7B	&	350.4968129	&	1.9257211	&	5	&	[L9]	&	6	&	120.5	$\pm$	24.6	&	138.6	$\pm$	24.7	&	10	&	4.9	&	200	&	\dots	\\	
\enddata
\tablerefs{(1) NSC DR2 \citep{nidever2021}; (2) UHS DR3 \citep{schneider2025}; (3) CatWISE2020 \citep{marocco2021}; (4) VHS \citep{mcmahon2013}; (5) UKIDSS \citep{lawrence2007}; (6) this work; (7) this work and \cite{best2018}; (8) unTimely detections \citep{meisner2023} (this work); (9) PS1 detections \citep{chambers2016} (this work); (10) NSC DR2 detections \citep{nidever2021} (this work)}
\tablenotetext{a}{Spectral types in square brackets are photometric estimates.}
\tablenotetext{b}{Separations in au are based on the photometric $J$-band distance estimates of the primaries in each system.}
\tablenotetext{c}{No $J$-band magnitude is available for CWISE J132836.50+635527.7A, so the W2 photometric distance (69.1 pc) was used to calculate the separation of this pair in au.}
\end{deluxetable*} 
\end{longrotatetable}

\section{Conclusion} 
\label{sec:conclusion}

This work presents some of the primary results of years worth of effort by thousands of volunteers working with the BYW project to discover new proper motion objects using WISE data.  This effort has resulted in over 3,000 new motion-confirmed L and T dwarf candidates.  Among these new discoveries, we have identified 28 likely new comovers to higher-mass stars and 9 potential wide ultracool pairs. Many of these systems are deserving of detailed follow-up observations and analyses. 

The UltracoolSheet\footnote{http://bit.ly/UltracoolSheet}, which is the most complete compilation of spectroscopically confirmed ultracool dwarfs available, lists a total of 2,930 objects with spectral types of L0 or later in its most recent version.  If confirmed, the $>$3,000 new motion candidates found through the BYW program would more than double the number of objects with L, T, and Y spectral types.  This catalog of sources provides a legacy dataset to initiate numerous new investigations, including color-outliers, cluster and moving group members, old low-metallicity objects, and other binary or multiple systems. 

\begin{acknowledgments}
This publication makes use of data products from the UKIRT Hemisphere Survey, which is a joint project of the United States Naval Observatory, the University of Hawaii Institute for Astronomy, the Cambridge University Cambridge Astronomy Survey Unit, and the University of Edinburgh Wide-Field Astronomy Unit (WFAU). The WFAU gratefully acknowledges support for this work from the Science and Technology Facilities Council (STFC) through ST/T002956/1 and previous grants. SD acknowledges support from a UK STFC grant (ST/X000982/1).  JF acknowledges NSF CAREER award 2238468 and NASA award \#80NSSC22K0491 for support of this work.

This work is based in part on data obtained as part of the UKIRT Infrared Deep Sky Survey. The UKIDSS project is defined in \cite{lawrence2007}. UKIDSS uses the UKIRT Wide Field Camera (WFCAM; \citealt{casali2007}) and a photometric system described in \cite{hewett2006}. The pipeline processing and science archive are described in \cite{irwin2004} and \cite{hambly2008}.

The Pan-STARRS1 Surveys (PS1) and the PS1 public science archive have been made possible through contributions by the Institute for Astronomy, the University of Hawaii, the Pan-STARRS Project Office, the Max-Planck Society and its participating institutes, the Max Planck Institute for Astronomy, Heidelberg and the Max Planck Institute for Extraterrestrial Physics, Garching, The Johns Hopkins University, Durham University, the University of Edinburgh, the Queen's University Belfast, the Harvard-Smithsonian Center for Astrophysics, the Las Cumbres Observatory Global Telescope Network Incorporated, the National Central University of Taiwan, the Space Telescope Science Institute, the National Aeronautics and Space Administration under Grant No. NNX08AR22G issued through the Planetary Science Division of the NASA Science Mission Directorate, the National Science Foundation Grant No. AST–1238877, the University of Maryland, Eotvos Lorand University (ELTE), the Los Alamos National Laboratory, and the Gordon and Betty Moore Foundation.

This publication makes use of data products from the Wide-field Infrared Survey Explorer, which is a joint project of the University of California, Los Angeles, and the Jet Propulsion Laboratory/California Institute of Technology, funded by the National Aeronautics and Space Administration. This publication also makes use of data products from NEOWISE, which is a project of the Jet Propulsion Laboratory/California Institute of Technology, funded by the Planetary Science Division of the National Aeronautics and Space Administration.

This project used data obtained with the Dark Energy Camera (DECam), which was constructed by the Dark Energy Survey (DES) collaboration. Funding for the DES Projects has been provided by the US Department of Energy, the U.S. National Science Foundation, the Ministry of Science and Education of Spain, the Science and Technology Facilities Council of the United Kingdom, the Higher Education Funding Council for England, the National Center for Supercomputing Applications at the University of Illinois at Urbana-Champaign, the Kavli Institute for Cosmological Physics at the University of Chicago, Center for Cosmology and Astro-Particle Physics at the Ohio State University, the Mitchell Institute for Fundamental Physics and Astronomy at Texas A\&M University, Financiadora de Estudos e Projetos, Funda{\c c}{\~a}o Carlos Chagas Filho de Amparo {\'a} Pesquisa do Estado do Rio de Janeiro, Conselho Nacional de Desenvolvimento Cient{\'i}fico e Tecnol{\'o}gico and the Minist{\'e}rio da Ci{\^e}ncia, Tecnologia e Inova{\c c}{\~ a}o, the Deutsche Forschungsgemeinschaft and the Collaborating Institutions in the Dark Energy Survey.  The Collaborating Institutions are Argonne National Laboratory, the University of California at Santa Cruz, the University of Cambridge, Centro de Investigaciones En{\'e}rgeticas, Medioambientales y Tecnol{\'o}gicas–Madrid, the University of Chicago, University College London, the DES-Brazil Consortium, the University of Edinburgh, the Eidgen{\"e}ssische Technische Hochschule (ETH) Z{\"u}rich, Fermi National Accelerator Laboratory, the University of Illinois at Urbana-Champaign, the Institut de Ci{\'e}ncies de l’Espai (IEEC/CSIC), the Institut de F{\'i}sica d’Altes Energies, Lawrence Berkeley National Laboratory, the Ludwig-Maximilians Universit{\"a}t M{\"u}nchen and the associated Excellence Cluster Universe, the University of Michigan, NSF NOIRLab, the University of Nottingham, the Ohio State University, the OzDES Membership Consortium, the University of Pennsylvania, the University of Portsmouth, SLAC National Accelerator Laboratory, Stanford University, the University of Sussex, and Texas A\&M University.

This work has made use of data from the European Space Agency (ESA) mission
{\it Gaia} (\url{https://www.cosmos.esa.int/gaia}), processed by the {\it Gaia}
Data Processing and Analysis Consortium (DPAC,
\url{https://www.cosmos.esa.int/web/gaia/dpac/consortium}). Funding for the DPAC
has been provided by national institutions, in particular the institutions
participating in the {\it Gaia} Multilateral Agreement.

This publication makes use of data products from the Two Micron All Sky Survey, which is a joint project of the University of Massachusetts and the Infrared Processing and Analysis Center/California Institute of Technology, funded by the National Aeronautics and Space Administration and the National Science Foundation.

This research uses services or data provided by the Astro Data Lab, which is part of the Community Science and Data Center (CSDC) Program of NSF NOIRLab. NOIRLab is operated by the Association of Universities for Research in Astronomy (AURA), Inc. under a cooperative agreement with the U.S. National Science Foundation.

This work has benefitted from The UltracoolSheet at http://bit.ly/UltracoolSheet, maintained by Will Best, Trent Dupuy, Michael Liu, Aniket Sanghi, Rob Siverd, and Zhoujian Zhang, and developed from compilations by \cite{dupuy2012}, \cite{dupuy2013}, \cite{liu2016}, \cite{best2018}, \cite{best2021}, \cite{sanghi2023}, and \cite{schneider2023}.

\end{acknowledgments}

\facilities{WISE, NEOWISE, UKIRT, ESO:VISTA, PS1, Blanco, Mayall, Gaia, IRSA, MAST, Astro Data Archive, Astro Data Lab}

\software{\texttt{CoMover} \citep{gagne2021}}

\bibliography{references1}{}
\bibliographystyle{aasjournal}

\end{document}